\documentclass{aastex}
\usepackage{emulateapj5}
\usepackage{apjfonts}
\usepackage{epsf}

\slugcomment{Accepted for Publication in The Astrophysical Journal Letters}

\shorttitle{Narrow iron K$\alpha$ lines in AGNs}
\shortauthors{Zhou \& Wang}

\newcommand{\ce}{\ifmmode {\cal E} \else ${\cal E}$\ \fi}
\newcommand{\kms}{\ifmmode {\rm km\ s}^{-1} \else km s$^{-1}$\ \fi}
\newcommand{\ergs}{\ifmmode {\rm erg\ s}^{-1} \else erg s$^{-1}$\ \fi}
\newcommand{\tes}{\ifmmode \tau_{\rm es} \else $\tau_{\rm es}$\ \fi}
\newcommand{\tk}{\ifmmode \tau_{\rm K} \else $\tau_{\rm K}$\ \fi}
\newcommand{\vfwhm}{\ifmmode V_{\mbox{\tiny FWHM}} \else
            $V_{\mbox{\tiny FWHM}}$\fi}
\newcommand{\msun}{\ifmmode M_{\odot} \else $M_{\odot}$\ \fi}
\newcommand{\afe}{\ifmmode {\mathcal A_{\rm Fe}} \else${\mathcal A_{\rm Fe}}$\ \fi}
\newcommand{\et}{et al.\ }

\newcommand{\nv}{N {\sc v}\ }

\newcommand{\civ}{C {\sc iv}\ }

\newcommand{\lb}{\ifmmode L_{\rm Bol} \else $L_{\rm Bol}$\ \fi}
\newcommand{\ledd}{\ifmmode L_{\rm Edd} \else $L_{\rm Edd}$\ \fi}
\newcommand{\lx}{\ifmmode L_{\rm 2-10keV} \else  $L_{\rm 2-10keV}$\ \fi}
\newcommand{\hb}{\ifmmode H\beta \else H$\beta$\ \fi}
\newcommand{\mbh}{\ifmmode M_{\rm BH}  \else $M_{\rm BH}$\ \fi}
\newcommand{\lv}{\ifmmode \lambda L_{\lambda}(5100\AA) \else $\lambda L_{\lambda}(5100\AA)$\ \fi}

\received{2004 August 2}
\begin{document}

\title{NARROW IRON K$\alpha$ LINES IN ACTIVE GALACTIC NUCLEI: EVOLVING POPULATIONS?}

\author{Xin-Lin Zhou\altaffilmark{1,2} and
Jian-Min Wang\altaffilmark{1,2,3}}

\altaffiltext{1}{Laboratory for High Energy Astrophysics,
       Institute of High Energy Physics, Chinese Academy of Sciences,
       Beijing 100039, P.R\@. China, wangjm@mail.ihep.ac.cn}
\altaffiltext{2}{Graduate School of Chinese Academy of Sciences, Beijing 100039, P.R China}
\altaffiltext{3}{to whom should correspond}

\begin{abstract}
We assemble a sample consisting of 66 active galactic nuclei
(AGNs) from literature and the {\em XMM-Newton} archive in order
to investigate the origin of the 6.4\,keV narrow iron K$\alpha$
line (NIKAL). The X-ray Baldwin effect of the NIKAL is confirmed
in this sample. We find the equivalent width ($EW$) of the NIKAL
is more strongly inversely correlated with Eddington
ratio ($\ce$) than the 2-10\,keV X-ray luminosity. Our sample favors 
the origin from the dusty torus and the X-ray Baldwin effect is caused 
by the changing opening angle of the dusty torus. The relation $EW-\ce$
can be derived from a toy model of the dusty torus. If the
unification scheme is valid in all AGNs, we can derive the Baldwin
effect from the fraction of type {\sc ii} AGNs to the total
population given by {\em Chandra} and {\em Hubble} deep surveys.
Thus the evolution of populations could be reflected by the
NIKAL's Baldwin effect.

\end{abstract}

\section{INTRODUCTION}

Recent observations
by {\em Chandra} and {\em XMM-Newton} have revealed that
a narrow iron K$\alpha$ line (NIKAL) at 6.4\,keV in the rest frame 
is a ubiquitous feature in AGN X-ray spectra. However, its origin 
is poorly understood.

There are two possible ways to test the origins of the NIKAL.
Firstly, the full-width-half-maximum (FWHM) of the line may
provide the motion of the emitting matter, indicating the location
of the radiation region in the potential of the black hole at the
center. The currently-available highest spectral resolution of
{\em Chandra} is $\sim$ 1860\,\kms (FWHM) at 6.4\,keV and {\em
XMM-Newton} has a resolution of $\sim 7200$\,\kms at the same
energies. There are a dozen of objects which have been resolved by
{\em Chandra} and {\em XMM}-Newton, e.g\@. $\sim
1780^{+1420}_{-1220}$ \kms in NGC 5548; $\sim 1700^{+410}_{-390}$
\kms in NGC 3783 (Yaqoob \et 2004). However, the current
observational data set up less constraint on the NIKAL's origin
since they cover a range of $10^3\sim 10^4$ \kms and the error
bars are quite large.

Secondly, the variabilities of the NIKAL are expected to apply to
test its origin. However, it appears rather constant and lacks the
response to the changes in the continuum flux, for example,
MCG-6-30-15 (Wilms \et 2001, Lee \et 2002), Mrk 6 (Immler \et
2003), NGC 4051 (Pounds \et 2004) and NGC 4593 (Reynolds \et
2004). In our present sample, only Mrk 841 shows a rapid variation
of the NIKAL with a timescale of 10 hours or less whereas the
X-ray continuum is constant (Petrucci \et 2002). The available
data implies a rather constant geometry of the NIKAL's emission
region.

Despite of the illusive properties of the NIKAL, there is an
anti-correlation between the equivalent width ($EW$) of the NIKAL
and X-ray continuum, namely, the X-ray Baldwin effect (Iwasawa \&
Taniguchi 1993, Nandra \et 1997). This is recently confirmed in
the {\em XMM-Newton} data by Page \et (2004, hereafter P04). They
argue that the origin of the NIKAL from BLR is not favored by: (1)
BLR could only account for the NIKAL with a $EW$ of less than
40\,eV (see also Yaqoob \et 2001); (2) the lack of the correlation
between the NIKAL and optical/UV parameters. However, the origin
of the NIKAL from a dusty torus has been explored by Krolik \&
Kallman (1987, hereafter KK87) and Krolik et al\@. (1994). If the
NIKAL indeed originates from the dusty torus, its subtended solid
angle is decisive to the observed $EW$ of the NIKAL. This implies
that the opening angle of the dusty torus could be indicated by
the $EW$ of the NIKAL. In turn, the evolution of the open angle of
the dusty torus could be reflected by the X-ray Baldwin effect.

In this Letter, we test the origin of the NIKAL by revisiting the
{\em XMM-Newton} data. We find that the NIKAL $EW$ much more
strongly depends on the Eddington ratio and the current data
agrees with an origin of evolving dusty torus.

\section{THE SAMPLE AND STATISTICAL PROPERTIES}
We collect all the AGNs with observations of {\em XMM-Newton} from
literature and find 66 objects listed in Tab.\,1. In order to
minimize systematic errors in the correlation analysis, we use
only the results obtained in P04 who give the NIKAL of 51 of 66
objects. We analyze the PN data of the additional 15 objects from
the {\em XMM-Newton} archive according to the model described in
P04. The sample consists of 16 narrow line Seyfert 1 galaxies
(NLS1), 18 broad line Seyfert 1 galaxies (BLS1), 32 quasars
including 14 radio-loud (RL) quasars. The redshift range of the
present sample is $z=0.002-3.4$, but only 7 objects have $z>0.5$.

\begin{figure*}[t]
\centerline{\includegraphics[angle=-90,width=16cm]{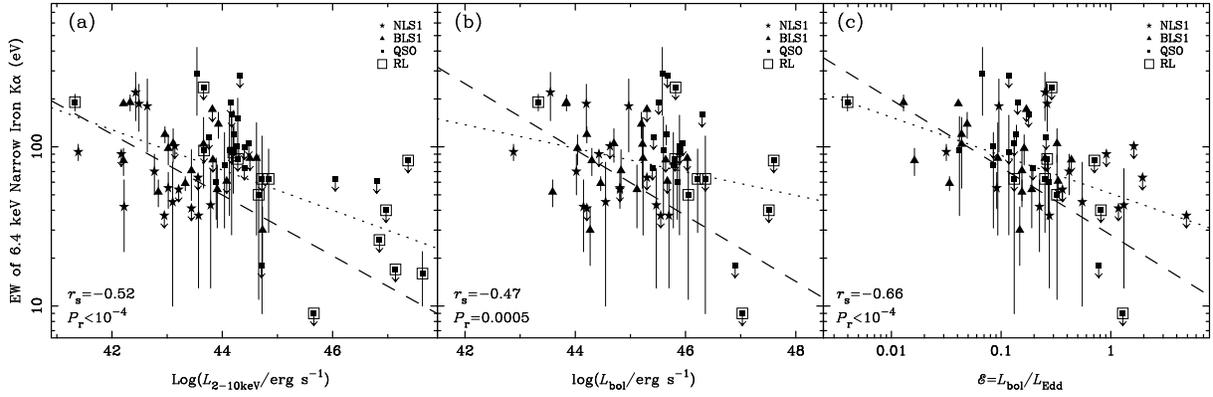}}
\caption{\footnotesize ({\em a}): the  X-ray Baldwin effect; ({\em
b}): the correlation of $EW-L_{\rm Bol}$; ({\em c}): the strong
anti-correlation of $EW$ with Eddington ratio ${\cal E}$. The
dashed lines indicates the relation for the entire dataset, dotted
line without upper limits. \label{fg-1}}
\end{figure*}

\def\hb{H$\beta$}
\def\eka{$EW_{\rm K\alpha}$}
\def\lbhm{$\log M_{\rm BH}$}
\def\lce{$\log {\cal E}$}
\def\llbol{$\log L_{\rm Bol}$}
\def\llx{$\log L_{\rm X}$}

\begin{table*}[t]
\begin{center}
\footnotesize
\centerline{\sc Table 1. The {\em XMM-Newton} sample of AGNs
(see the complete electronic version of the Table)}
\vglue 0.1cm
\begin{tabular}{lllcccclrlllc}\hline \hline
Name        &$z$          & Type &\llbol&   &\hb  &\lbhm  &    &\lce     &Ref.   &\llx    &\eka  &Ref. \\
(1)         & (2)    & (3)  & (4)  &(5)& (6) &  (7)  & (8)& (9)     & (10)  & (11)   & (12)      & (13) \\\hline
1H 0707-495 & 0.0411 & NLS1 &44.43 & R& 1050 &  6.37 & R  & $-0.04$ & 11,13 & 42.17  & $<90$     & 1 \\
I Zw 1$^{b}$& 0.0611 & NLS1 &45.47 & R& 1240 &  7.25 & R  & $ 0.12$ & 3,13  & 43.79  & $34\pm30$ & 2\\ \hline
\end{tabular}
\end{center}
\end{table*}

The black-hole masses ($M_{\rm BH}$) are directly taken from the
results of the reverberation mapping measurements of Peterson et
al\@. (2004), or estimated from the empirical reverberation
relation (Kaspi et al\@. 2000; hereafter K00), $M_{\rm BH}/M_{\odot}=4.82\times
10^6\left[\lambda L_{\lambda, 44}(5100{\rm
\AA})\right]^{0.7}V_3^2$, where $\lambda L_{\lambda, 44}(5100{\rm
\AA})=\lambda L_{\lambda}(5100{\rm \AA})/10^{44}\ergs$ and
$V_3={\rm FWHM(H\beta)}/10^3\kms$. The values of
$L_{\lambda}(5100{\rm \AA})$ are taken from  Vestergaard (2002)
and Grupe \et (2004), or calculated from the absolute magnitude in
the $B$ band of $M_{\rm B}$ given by Marziani \et (2003) and
V\'eron-Cetty \et (2003) through an extrapolation of a power-law
spectrum as $F_{\rm \nu} \propto \nu^{-0.5}$. Additionally, 4
objects in the present sample are estimated via different methods,
including 2 high-redshift quasars, 2 Seyfert 1.9. Only 57 of 66
objects have enough data to estimate the BH masses. We estimate
bolometric luminosity for the Eddington ratio from $ \lb= 9\lv$
(K00) and from published papers if it is given. We then have 

\begin{center}
\footnotesize
\centerline{Table 2. The NIKAL $EW$ correlation analysis}
\label{tbl-2}
\vskip 0.1cm
\begin{tabular}{llcrcccccc}\hline \hline
   $X$    & data   &$r_{\rm s}$   & $P_{\rm r}$  & $a$ & $b$ \\ \hline
$L_{\rm X}$& All            & $-0.52$  &  $<10^{-4}$        &$-0.19\pm0.04$ &$10.10\pm1.49$\\
           & RQ only        & $-0.34$  &  $1.6\times10^{-2}$&$-0.15\pm0.05$ &$8.51\pm2.30$ \\
           & Real           & $-0.38$  &  $2.8\times10^{-2}$&$-0.13\pm0.03$ &$7.42\pm1.47$ \\ \hline
\lb        & All            & $-0.47$  &  $5.0\times10^{-4}$&$-0.21\pm0.05$ &$11.10\pm2.20$\\
           & RQ only        & $-0.36$  &  $1.1\times10^{-2}$&$-0.15\pm0.06$ &$8.64 \pm2.58$\\
           & Real           & $-0.28$  &  $0.12$            &$-0.07\pm0.05$ &$5.25\pm2.25$\\ \hline
\ce        & All            & $-0.66$  &  $<10^{-4}$        &$-0.42\pm0.07$ &$1.45\pm0.08$\\
           & RQ only        & $-0.61$  &  $<10^{-4}$        &$-0.38\pm0.08$ &$1.52\pm0.08$ \\q
           & Real           & $-0.49$  &  $6.4\times10^{-3}$&$-0.24\pm0.07$ &$1.71\pm0.08$ \\
 \hline
\end{tabular}
\vskip 0.1cm
\parbox{3.45in}
{\baselineskip 9pt
\indent
{ Note:--
$\log EW =a\log X+b$\\
$r_{\rm s}$ is the Spearman's coefficient; $P_{\rm r}$ is the null probability.
}}
\end{center}
\normalsize

\noindent the
Eddington ratio $\ce \equiv \lb/\ledd$, where \lb is the
bolometric luminosity and \ledd is the Eddington luminosity.
Throughout this paper, we use $H_0$ = 75~km~s$^{-1}$~Mpc$^{-1}$
and $q_0=0.5$.

Since the present sample includes upper limit data, we apply the
Spearman's rank statistic and parameter EM
(Expectation-Maximization) algorithm constructed by {\sc ASURV}
(Astronomy Survival Analysis; Isobe \et 1986) package to the
complete dataset for the correlation analysis. Table 2 gives the
correlation analysis of the equivalent width with $L_X$, $L_{\rm
Bol}$ and ${\cal E}$. Fig\@. 1 presents the plots of these
correlations.

The X-ray Baldwin effect is confirmed by the present sample which
is larger compared to P04. We find $EW\propto L_{\rm X}^{-0.19\pm
0.04}$ for the entire data. We have 16 NLS1s in our sample, which
are frequently variable in X-ray. We thus explore the correlation
between $EW$ and the bolometric luminosity $L_{\rm Bol}$. We find
$EW-L_{\rm Bol}$ is not better than $EW-L_{\rm X}$ as shown in
Fig. 1{\em b} and Table 2. Interestingly, we find a much stronger
correlation of $EW\propto \ce^{-0.42\pm 0.07}$ for the entire
sample as shown by Fig\@. 1{\em c}, indicating that the
correlation of $EW-{\cal E}$ is more fundamental than $EW-L_{\rm
X}$. What does drive the correlations?

\section{THE ORIGIN OF THE X-RAY BALDWIN EFFECT}
The equivalent width of Fe K$\alpha$ line from the illuminated medium
is given by KK87,
\begin{equation}
\label{ew}
EW \approx \frac{E_{\rm K\alpha}}{3+\Gamma}Y
\left(\frac{\Delta \Omega}{4\pi}\right)\tk,
\end{equation}
where $\tk \approx 0.7f(\Xi)\afe \tes$. Here is $E_{\rm
K\alpha}=6.4~{\rm keV}$; $\Gamma$ is the photon index of 2-10 keV
band; $Y$ is the photon production; $\Delta \Omega$ is the solid
angle subtended by the illuminated matter; \tk is the Fe K-edge
optical depth; $\Xi$ is the ionization parameter, $f(\Xi)$ is a
weak function of $\Xi$, normalized as $f(30)=1$ and $f$ approaches
1.7 in the limit of small $\Xi$; \afe is the iron abundance in
units of solar abundance; \tes is the electron scattering optical
depth.

P04 excludes the dependence of the NIKAL $EW$ on the index
$\Gamma$. Since we pay our attention to the NIKAL, the ionization
parameter $\Xi$ should be small enough and $Y$ keeps a constant.
The dependence of the NIKAL on $\Xi$ and $Y$ should be ruled out.
The metallicity indicator (e.g., \nv/\civ) measured from the BLR
in quasars shows that the metallicity increases with luminosity
(Hamann \& Ferland 1993), and more strongly with the Eddington
ratio (Shemmer \et 2004). Such a dependence on the metallicity
leads to an increases of the NIKAL $EW$ with the Eddington ratio.
This is just opposite to the present correlation shown by Fig
1{\em c}. Risaliti \et (1999) find that the column density does
not relate to the hard X-ray luminosity. This means that the Fe K
edge depth can not result in the Baldwin effect. Iwasawa \&
Tanuguchi (1993) point out that the X-ray Baldwin effect may be
caused by a changing covering factor similar to the optical
Baldwin effect (Mushotzky \& Ferland 1984), but they do not give
more details.

\subsection{Origin from Outer Region of Disk}
The covering factor $\Delta \Omega/ 4 \pi$ in Eq.\ref{ew} is
crucial to the $EW$ of the NIKAL, implying the importance of
geometry of the emission region. In the followings, we explore the
probability that the NIKAL originates from the outer region of the
accretion disk and the torus.

\subsubsection{The outer region of the standard accretion disk}
The solid angle subtended by the disk height can be derived from
the standard accretion disk (Shakura \& Sunyaev 1973)
\begin{equation}
\Delta \Omega \propto \frac{H}{R}\propto \left\{ \begin{array}{ll}
{\cal E}^{3/20}  & \mbox{($P_{\rm gas} \gg P_{\rm rad}$}),\\
{\cal E} & \mbox{($P_{\rm rad} \gg P_{\rm gas}$),} \end {array} \right.
\end{equation}
where $H$ is the half height of the disk at radius $R$ from the center, 
$P_{\rm gas}$ and $P_{\rm rad}$ are the gas and radiation pressure, 
respectively. This directly indicates that the $EW$ of iron K$\alpha$ 
line increases with the Eddington ratio, as $EW\propto \ce^{3/20}$ or 
$EW \propto \ce$. The correlation shown in Fig 1{\em c} rules out the 
origin of the NIKAL from outer region of the accretion disks.

For low Eddington ratios, the advection-dominated accretion flows (ADAF) 
can automatically switch on the standard disks at a radius larger than 
10$R_g$, where $R_g=2GM_{\rm BH}/c^2$ (Lu et al\@. 2004). For the slim 
disks, Chen \& Wang (2004) show that they have structures similar
to the standard disks beyond the photon trapping radius $R_{\rm tr}$, 
where $R_{\rm tr}/R_g=720\left(\dot{m}/50\right)$, 
$\dot{m}=\eta \dot{M}c^2/L_{\rm Edd}$ and $\eta=0.1$ (Wang \& Zhou 1999, 
Wang \& Netzer 2003). Thus our discussions (eq\@. 2) are available for 
both kinds of the objects with low and high Eddington ratios, ruling out 
the possible origin from the outer region.

\subsubsection{The self-gravity disk}
The outer parts of the standard accretion disks tend to be
self-gravity-dominated, Goodman et al\@. (2003) show
\begin{equation}
\Delta \Omega \propto  \frac{H}{R} \propto \left\{ \begin{array}{ll}
         Q^{-5/3}\ce^{1/3}  & \mbox{($\alpha \propto P_{\rm tot}$),}\\
      Q^{-6/7} \ce^{2/3} & \mbox{($\alpha \propto P_{\rm gas}$),}\end{array} \right.
\end{equation}
where $Q$ is the Toomre stability parameter for Keplerian rotation,
$\alpha$ is the viscosity and $P_{\rm tot}$
is the total pressure. We have $EW\propto \ce^{1/3}$ or $EW\propto \ce^{2/3}$.
This shows that the NIKAL from a self-gravity-dominated disk will
strengthen with $\ce$. This origin is also ruled out by Fig 1{\em c}.

\subsection{The dusty torus}
The mid-infrared interferometry telescope reveals a geometrically and optically thick
torus in the well-known Seyfert 2 galaxy NGC 1068 (Jaffe \et 2004, hereafter).  The
torus has a thickness of $H/R>0.6$ and a temperature of 800-1000 K for the inner shell,
300 K for outer shell, where $H$ is the half height of torus at the distance $R$ from
its central engine. Such an unusual result provides direct support for the unified model
(Antonucci 1993). The covering factor plays a key role in the unification scheme, but it
is poorly understood. We employ the torus model to understand the X-ray Baldwin effect
from two ways.

\subsubsection{Covering factors of the tori and the accretion rates}
Krolik \& Begelman (1988) suggest that the obscuring matter takes the form of compressed
molecular clouds. They derive the covering factor of the torus
based on a toy model for cloud size distribution,
\begin{equation}
{\cal C}=\frac{1}{4}\left(\frac{\alpha}{\gamma}\right)
                    \left(\frac{V_{\rm orb}}{\Delta V}\right)
\end{equation}
where the two parameters $\alpha\sim 1$ and $\gamma\sim 1$, $V_{\rm orb}$ is the orbital
velocity of the clouds at radius $R$ and
$\Delta V^2=\Delta V_z^2+\Delta V_r^2+\Delta V_{\theta}^2\approx 3\Delta V_z^2$
is the square of the random velocity. The collisions due to the random motions of the
clouds lead to an inflow to the central black hole with a mass rate of
\begin{equation}
\dot{M}=1.5~V_{200}R_{\rm pc}{\cal C}^{-2}~M_{\odot}{\rm yr^{-1}},
\end{equation}
where $V_{200}=V_{\rm orb}/200~{\rm km~s^{-1}}$ and the radius of the inner edge of the torus
$R_{\rm pc}=R_t/1{\rm pc}$. We thus have a
relation of ${\cal C}\propto r_t^{1/4}{\cal E}^{-1/2}$, where $r_t=R_t/R_g$. The covering factor
is insensitive to the inner radius ($r_t$) of the torus, but very sensitive to the Eddington ratio.
We then have  $EW\propto {\cal E}^{-0.5}$.
This quite nicely matches the correlation $EW\propto {\cal E}^{-0.42\pm 0.07}$ (in Table 2).
This consistence shows an intrinsic relation between the accretion disk and the torus.

\subsubsection{Type {\sc ii} AGN Populations}
On the other hand, the covering factor of the torus can be derived from the
statistics of the populations of AGNs in deep sky surveys if the unification
scheme works for all AGNs.
The deep survey data from {\it Chandra}, {\em Hubble} and others show that the population
of type {\sc ii} AGNs is decreasing with X-ray luminosity (Ueda \et 2003, Steffen \et 2003,
Hasinger 2003). This census indicates the changing of the opening
angle in different individuals.
Using the least square method, we fit the data in Fig 6{\em b} in Hasinger (2003) and obtain
\begin{equation}
\label{PII}
\log P_{\rm II} = (3.38 \pm 0.48) - (0.08 \pm 0.02) \log \lx
\end{equation}
This relation shows that type {\sc ii} quasars will be rare among
luminous quasars. This is consistent with the fact that the NIKAL
disappears in luminous quasars. If the half-open angle of torus is
$\theta$, the solid angle of torus subtending the central engine
is $\Delta \Omega/4\pi = \cos \theta=P_{\rm II}$. This relation is
based on assumption that the orientation of the torus is random in
the sky. %It allows us to set up the connection between the opening
%angle of the dusty torus and hard X-ray luminosity, and hence the
%X-ray Baldwin effect.
Inserting eq\@.
(6) into (1) with the help of $\Delta \Omega/4\pi = \cos
\theta=P_{\rm II}$ and $\Gamma=1.7$, we have
\begin{equation}
\label{TE} \log EW = (6.20 \pm 0.48) - (0.08 \pm 0.02) \log \lx.
\end{equation}
We take $\tau_{\rm K}=1$, otherwise $\tau_{\rm K}$ is too large
for the fluorescence photons to escape whereas $\tau_{\rm K}$ is
too small to produce the NIKAL (Krolik et al\@. 1994; Levenson et
al\@. 2002).

It is very interesting to note that this relation is consistent
with the Baldwin effect described in Table 2. This suggests an
{\em intrinsic} relation between the NIKAL Baldwin effect and
evolution of the dusty torus. The fraction of type {\sc ii} AGNs
in the deep surveys indicates that the increasing X-ray luminosity
enlarges the opening angle of the dusty torus and lowers the
covering factor of the dusty torus. Quantitative calculations of
this process still lack in the literatures. In addition, the
collisions among the molecular clouds result in an evolution of
the torus covering factors and non-thermal emissions detectable in
radio to sub-GeV $\gamma$-ray bands (Wang 2004). Though the
explanation of $L_X-P_{\rm II}$ relation remains open, the canonic
evolutionary model of the dusty torus can be reflected by the
X-ray Baldwin effect.

We should emphasize that the derived Baldwin effect is based on
the deep survey results of {\em Chandra}, {\em Hubble}, etc.,
which is different from ours. Hence Eq. (7) is not
self-consistent. Though the fraction $P_{\rm II}$ also depends on
redshift (Steffen \et 2003), most of the present sample are
low-redshift AGNs, the dependence of $P_{\rm II}$ on the redshift
can be ignored. If the survey data is available for iron K$\alpha$
line analysis, it would be interesting to confirm the present
results using the objects in the deep surveys. We would like to
stress that the connection of the opening angle to the fraction of
type {\sc ii} ANGs is not firmly sure since the narrow line region
in high luminosity quasars may be different from that in Seyfert
galaxies (Netzer et al\@. 2004) and the properties of the highly
obscured AGN are not clear (Maiolino et al\@. 2003). With these
uncertainties, the present relation, as a preliminary element,
between the Baldwin effect and the evolution of the dusty torus is
still robust.

Finally, we like to point out that the presently available data shows
only statistically that the origin of the NIKAL is mainly from the dusty 
torus, however, it might have components from BLR or outer disk region as 
well. Actually, the scatter in the plot of $EW-L_X$ and $EW-\ce$ may be 
large because of these components of the observed NIKAL $EW$.
The current data does not allow to subtract the component of the NIKAL 
which originates from BLR or disk outer region.
Future observations of {\em Astro-E2}, {\em XEUS} and {\em Constellation-X} 
may resolve the different narrow components so that the correlation 
${EW-\cal E}$ will be improved.

\section{CONCLUSIONS}
In this paper, we show that the equivalent width of the neutral narrow iron 
K$\alpha$ line in AGN strongly depends on the Eddington ratio.
We exclude the origin from the outer region of the standard accretion disk 
or self-gravity-dominated disk. The current data statistically supports 
the origin of this narrow component from the dusty torus. The strong 
anti-correlation of $EW-\ce$ provides a robust probe to the covering factor 
of the dusty torus. The most promising explanation of the X-ray Baldwin 
effect is that it is caused by the evolution of the dusty torus with the 
hard X-ray luminosity. This is consistent with the population
evolution found in the deep surveys of the {\em Chandra} and {\em HST}.

\acknowledgements{The authors are very grateful to an anonymous
referee for the very helpful comments improving the paper. T.
Rauch is thanked for careful reading of the manuscript. J.M.W\@.
thanks the supports from a Grant for Distinguished Young Scientist
from NSFC, NSFC-10233030, the Hundred Talent Program of CAS and
the 973 project.}

\end{document}

\begin{deluxetable}{lllcccclrlllc}
\tabletypesize{\scriptsize}
\tablewidth{0pt}
\tablecaption{The {\em XMM-Newton} sample of AGNs
\label{tbl-1}}
\tablehead{
\colhead{Name}               &\colhead{$z$}     &\colhead{Type}      & \colhead{$\log\lb$}       &
\colhead{ }                  &\colhead{FWHM}    &\colhead{$\log\mbh$}   & 
\colhead{ }                  &\colhead{$\log\ce$}  &\colhead{Ref.}     & $\log L_{\rm X}$                 &
\colhead{$EW_{\rm K\alpha}$} &\colhead{Ref.} \\
\colhead{ }                  &\colhead{ }       &\colhead{ }        & \colhead{}          & 
\colhead{ }                  &\colhead{}     &\colhead{}     &
\colhead{ }                  &\colhead{}    & \colhead{ }       & \colhead{}  & 
\colhead{ }                  &\colhead{}    \\
\colhead{(1)}                &\colhead{(2)}     &\colhead{(3)}      &\colhead{(4)}           & 
\colhead{(5)}                &\colhead{(6)}     &\colhead{(7)}      &\colhead{(8)}           & 
\colhead{(9)}                &\colhead{(10)}    &\colhead{(11)}     &\colhead{(12)}          & 
\colhead{(13)}}
\startdata

1H 0707-495           & 0.0411      & NLS1 &44.43 & R& 1050 &  6.37 & R  & $-0.04$ & 11,13 & 42.17  & $<90$        & 1 \\
I Zw 1$^{b}$          & 0.0611      & NLS1 &45.47 & R& 1240 &  7.25 & R  & $ 0.12$ & 3,13  & 43.79  & $34\pm30$    & 2\\
Akn 564               & 0.0247      & NLS1 &44.21 & F&  720 &  6.04 & R  & $ 0.07$ & 11,13 & 43.44  & $<41$        & 2\\
IRAS 13224-3809       & 0.0667      & NLS1 &45.55 & F&  650 &  6.76 & R  & $ 0.69$ & 11,13 & 42.95  &$<37$         & 2\\
MCG-5-23-16$^{a}$     & 0.0083      & NLS1 &44.81 & R&  209 &  7.85 & N  & $-1.04$ & 7,12  & 43.02  & $55\pm19$    & 2\\   
Mrk 335$^{a}$         & 0.0258      & NLS1 &44.81 & R& 1629 &  7.15 & M  & $-0.44$ & 4     & 43.21  &$<54$        & 1 \\
Mrk 359               & 0.0174      & NLS1 &43.55 & F&  480 &  6.05 & R  & $-0.60$ & 7,8   & 42.43  & $220\pm74$  & 1\\  
Mrk 493               & 0.0313      & NLS1 & 44.63& F&  800 &  6.32 & R  & $0.21$  & 14    & 43.14  &$<101$       & 1\\    
Mrk 766$^{a}$         & 0.0129      & NLS1 &44.55 & F& 1360 &  6.60 & R  & $-0.26$ & 15,14 & 43.10  &$45\pm35$    & 1 \\ 
Mrk 896               & 0.0264      & NLS1 &44.97 & F& 1135 &  6.79 & R  & $-1.02$ & 10,15 & 42.64  & $180\pm87$  & 1 \\ 
Mrk 1044              & 0.0165      & NLS1 & 44.20& F& 1310 &  6.68 & R  & $-0.58$ & 14,9  & 42.49  & $186\pm61$  & 1 \\ 
NGC 4051              & 0.0023      & NLS1 & 42.88& R& 1072 &  6.28 & M  & $-1.50$ & 4     & 41.39  & $93\pm11$   & 2\\
NGC 5506              & 0.0062      & NLS1 & 44.02& R& 1850 &  6.30 & R  & $-0.38$ & 7,12  & 42.77  & $70\pm20$   & 1 \\
NGC 7314$^{b}$        & 0.0047      & NLS1 & 44.15& R&  170 &  6.70 & G  & $-0.65$ & 5,18  & 42.22  & $42\pm20$   & 2 \\ 
PG 1211+143$^{a}$     & 0.0809      & NLS1 & 45.70& R& 1317 &  8.16 & M  & $-0.56$ & 4     & 43.57  & $37\pm24$   & 1\\
Ton S180              & 0.0620      & NLS1 & 45.30& R&  980 &  6.91 & R  & $ 0.29$ & 13,14 & 43.56  &$<64$        & 1 \\ 
1H 0419-577           & 0.1040      & BLS1 & 46.03& F& 3460 &  8.53 & R  & $-0.60$ & 14,16 & 44.50  &$<85 $       & 1 \\
3C 120                & 0.0330      & BLS1 & 45.12& R& 2205 &  7.74 & M  & $-0.72$ & 4     & 43.91  &$54\pm24$    & 2\\
ESO 198-G24           & 0.0455      & BLS1 & 45.22& F& 6400 &  8.48 & R  & $-1.36$ & 23,7  & 43.66  & $104\pm49$  & 1\\
Fairall 9$^{a}$       & 0.0470      & BLS1 & 45.20& R& 6901 &  8.41 & M  & $-1.31$ & 4     & 43.93  & $139\pm26$ & 1 \\
IC 4329A              & 0.0161      & BLS1 & 44.27& R& 6431 &  7.00 & M  & $-0.83$ & 4     & 44.73  & $30\pm12$  & 1 \\
MCG-6-30-15$^{a}$     & 0.0077      & BLS1 &43.59 & F& 2400 &  6.30 & R  & $-0.81$ & 13,23 & 42.84  & $52\pm10$  & 1 \\ 
Mrk 279               & 0.0305      & BLS1 & 44.83& R& 3385 &  7.54 & M  & $-0.81$ & 4     & 43.44  & $71\pm25$  & 2\\
Mrk 509$^{a}$         & 0.0344      & BLS1 & 45.23& R& 2715 &  8.16 & M  & $-1.03$ & 4     & 44.62  &$85\pm57$   & 1 \\ 
Mrk 841               & 0.0364      & BLS1 & 45.64& F& 3470 &  7.90 & R  & $-0.36$ & 3,14  & 43.83  &  $<83$     & 1 \\  
Mrk 926$^{a}$         & 0.0473      & BLS1 & 45.67& F& 6300 &  8.05 & R  & $-0.48$ & 8,15  & 44.08  &  $<61$     & 1 \\
NGC 3516$^{a}{^{b}}$  & 0.0088      & BLS1 & 43.84& R& 3353 &  7.63 & M  & $-1.89$ & 4     & 42.33  & $196\pm22$ & 1\\
NGC 3783              & 0.0097      & BLS1 & 44.21& R& 3093 &  7.47 & M  & $-1.36$ & 4     & 42.96  &$120\pm14$ &2\\   
NGC 4151              & 0.0033      & BLS1 & 43.83& R& 4248 &  7.12 & M  & $-1.39$ & 4     & 42.21  &$187\pm3$    & 1\\ 
NGC 4593              & 0.0090      & BLS1 & 44.04& R& 3769 &  6.73 & M  & $-0.79$ & 4     & 43.02  & $98\pm21$   & 1\\
NGC 5548              & 0.0172      & BLS1 & 44.46& R& 4044 &  7.83 & M  & $-1.47$ & 4     & 43.33  & $59\pm6$    & 1\\
NGC 7213$^{b}$        & 0.0060      & BLS1 & 44.30& I& 3200 &  7.99 & R  & $-1.79$ & 11,17 & 42.21  & $82\pm16$   &2\\
NGC 7469              & 0.0163      & BLS1 & 44.70& R& 2169 &  7.09 & M  & $-0.49$ & 4     & 43.11  & $105\pm25$  &2\\
PG 0844+349$^{a}$     & 0.0640      & BLS1 & 45.30& R& 2148 &  7.97 & M  & $-0.77$ & 4     & 43.82  &$<172$       &1 \\  
3C 273                & 0.1583      & RLQ  & 47.03& R& 3520 &  8.82 & R  & $ 0.11$ & 3     & 45.66  &$<9$         &1 \\ 
III Zw 2              & 0.0898      & RLQ  & 45.81& R& 3980 &  8.29 & R  & $-0.58$ & 3,11  & 44.28  &  $<84$      &1 \\  
B2 1028+31$^{a}$      & 0.1782      & RLQ  & -    & -&  -   &  -    & -  & -       &  -    & 44.20  & $88\pm41$   &1 \\ 
B2 1128+31$^{a}$      & 0.2890      & RLQ  & 46.05& R& 3400 &  8.44 & R  & $-0.49$ & 15,11 & 44.66  & $50\pm39$   &1 \\ 
B2 1721+34            & 0.2060      & RLQ  & 46.22& R& 3450 &  8.72 & R  & $-0.60$ & 15,11 & 44.84  & $63\pm34$   &1 \\
HE 1029-1401          & 0.0860      & RQQ  & 45.95& R& 7400 &  8.73 & R  & $-0.88$ & 15,11 & 44.48  & $<105$      &1 \\
MR 2251-178           & 0.0640      & RQQ  & 45.40& R& 4800 &  8.01 & R  & $-0.71$ & 6,15  & 44.40  &$ <74$       &1 \\ 
Mrk 205$^{a}$         & 0.0708      & RQQ  & 45.85& R& 3150 &  8.32 & R  & $-0.57$ & 6,15  & 43.89  & $60\pm25$   &1 \\ 
Mrk 304               & 0.0658      & RQQ  & 45.42& R& 6170 &  7.91 & R  & $-0.59$ & 11,24 & 43.76  & $<115$      &1 \\
Mrk 876               & 0.1290      & RQQ  & 45.60& F& 8450 &  8.88 & R  & $-1.38$ & 3,14  & 44.13  &$96\pm59$    &1 \\ 
Mrk 1383              & 0.0865      & RQQ  & 45.78& R& 5940 &  8.75 & R  & $-1.07$ & 3,11  & 44.04  &$ 77\pm46$   &1 \\ 
NGC 3227              & 0.0039      & RLQ  & 43.33& R& 5138 &  7.63 & M  & $-2.40$ & 4     & 41.33  &$191\pm23$   &1 \\ 
PB 05062              & 1.7700      & RQQ  &  -   & -& -    &  -    & -  &   -     &  -    & 46.80  & $<61$       &1 \\   
PDS 456               & 0.1840      & RQQ  & 46.90& R& 3010 &  8.91 & R  & $-0.11$ & 6,22  & 44.71  &$<18$        &1 \\     
PG 0804+761$^{a}$     & 0.1000      & RQQ  & 45.89& R& 2012 &  8.84 & M  & $-1.07$ & 4     & 44.26  &$<101$       &1 \\
PG 0947+396           & 0.2060      & RQQ  & 45.89& R& 4830 &  8.72 & R  & $-0.93$ & 21    & 44.17  & 93$\pm$65   &2\\
PG 1048+342           & 0.1670      & RQQ  & 45.51& R& 3890 &  8.26 & R  & $-0.85$ & 11,15 & 44.15  &  $<191$     &1 \\
PG 1114+445           & 0.1440      & RQQ  & 45.65& R& 4570 &  8.41 & R  & $-0.86$ & 3,6   & 44.21  & $120\pm17$  &1 \\  
PG 1309+355$^a$       & 0.1840      & RLQ  & 45.82& R& 2940 &  8.26 & R  & $-0.54$ & 21    & 43.66  & $<236$      &2 \\
PG 1411+442           & 0.0896      & RQQ  & 45.58& R& 2398 &  8.65 & M  & $-1.17$ & 4     & 43.54  &$290\pm132$  &2 \\
PG 1512+370           & 0.3707      & RLQ  & 46.36& R& 6810 & 9.14  & R  & $-0.88$ & 21    & 44.72  &63$\pm$54    &2 \\  
PG 1626+554           & 0.1330      & RQQ  & 45.67& F& 4490 & 8.50  & R  & $-0.93$ & 14,21 & 44.32  & $<281$      &1 \\
PG 1634+706           & 1.3340      & RQQ  & -    & -&  -   & -     & -  &  -      &  -    & 46.05  & $<63$       &1\\
PKS 0438-43           & 2.8630      & RLQ  & -    & -& -    & -     & -  &  -      &  -    & 46.85  & $<26$       &1\\       
PKS 0537-286          & 3.1040      & RLQ  & 47.60& R&12130$^c$&9.65& S  & $-0.15$ & 6,20  & 47.37  &$<82$        &1 \\
PKS 1637-77           & 0.0427      & RLQ  & -    & -&  -   &  -    & -  & -       &  -    & 43.67  & $<95$       &1 \\ 
PKS 2149-306          & 2.3450      & RLQ  & 47.51& R&6400$^c$&9.50 & S  & $-0.09$ & 6,20  & 46.97  & $<40$       &1 \\
Q 0056-363$^{a}$      & 0.1620      & RQQ  & 46.30& R& 4550  & 8.95 & R  & $-0.75$ & 14,11 & 44.17  & $<159$      &1 \\  
Q 0144-3938           & 0.2440      & RQQ  & -    &- & -     &  -   & -  &  -      &  -    & 44.28  & $152\pm50$  &1 \\
S5 0014+81            & 3.3660      & RLQ  & -    &- & -     &  -   & -  &  -      &  -    & 47.14  & $<17$       &1 \\
S5 0836+71            & 2.1720      & RLQ  & -    &- & -     &  -   & -  &  -      &  -    & 47.63  & $16\pm6$    &1 \\ 
UM 269                & 0.3084      & RQQ  & -    &- & -     &  -   & -  &  -      &  -    & 44.40  & $< 99$      &1 \\
    
\enddata
\tablecomments {(1): source name, (2): redshift, (3): type, (4): bolometric luminosity 
($\ergs$), (5): Note on method for estimates of bolometric luminosity. R: relation of $\lb=9\lv$ 
(K00, Netzer 2003); F: spectral energy distribution fitting; I: flux integration, (6): FWHM 
\hb $(\kms)$, (7): black-hole mass ($\msun$), (8):  note on method for estimates of black-hole 
masses, M: reverberation mapping measurement; R: the empirical reverberation relation; 
S: Single-epoch mass determination method, 
$\mbh=1.6\times10^6~(\lambda L_{\lambda,44}(1350\AA))^{0.7}(\vfwhm{_{\rm{\civ,3}}})^2$ $\msun$
(V02); N: relation for narrow line region size, $R_{\rm NLR}=1.15L_{\rm{\hb,42}}^{0.49}$ kpc 
(Netzer \et 2004); V: measurements of the central velocity gradient, taken from Schulz \et 
(1994), (9): Eddington ratio, $\ce=\lb/(1.26\times10^{38}\mbh)$, (10): reference for optical 
data, (11): the X-ray luminosity in 
the rest frame $2-10$ keV band ($\ergs$), (12): the rest frame Equivalent Width of neutral 
narrow Fe K$\alpha$ line at 90\% error confidence(eV). The {\em XMM-Newton} PN data only are 
used. As described in P04, the spectral fittings are performed via a model of a simple power 
law over the rest frame band of 2 - 10 keV modified by the Fe K edge. If there was evidence
 for the emission of broad line, a broad Gaussian model is included. Then a narrow Gaussian
(the intrinsic width is fixed at 10 eV ) is added for the EW measurements. 
The neutral reflection component ({\sc pexrav} in XSPEC, see Magdziarz \& Zdziarski 1995) is 
also tried for each spectrum. (13): reference for X-ray data}

\tablerefs{1: Page \et(2004); 2: this work; 3: Vestergaard \et(2002); 4: Peterson 
\et(2004); 5: Schulz \et(1994); 6: Reeves \& Turner (2000)  7: Whittle 
(1992); 8: Osterbrock \et(1982); 9: Rafanelli \et(1983); 10: Veilleux (1991); 
11: V\'{e}ron-Cetty \et(2003); 12: Wilson \et(1985); 13: Turner \et(1999);
14: Grupe \et(2004); 15: Marziani \et(2003);
16: Guainazzi \et(1998) 17: Woo \& Urry (2002); 18: Padovani \et(1988)
19: Vaughan \et(1999); 20: Wilkes (1986); 21: Porquet \et(2004); 22: Reeves \et(2000a); 
23: Winkler (1992); 24: Condon \et(1985)}
\tablenotetext{a}{Broad line is statistically required in the rest frame band of $2-10$ keV}
\tablenotetext{b}{Multiple narrow Gaussians in the rest-frame band of $2-10$ keV}
\tablenotetext{c}{Full width at half maximum of the \civ line (1549 $\AA$)}

\end{deluxetable}

\newpage